\documentclass[preprint,final,5p,times,twocolumn]{elsarticle}
\usepackage{rotating,color,subfigure}
\usepackage{alphalph}
\usepackage[T1]{fontenc}
\journal{Physics Letters B}

\usepackage{notoccite}
\usepackage{appendix}
\usepackage{physics}
\usepackage{mathrsfs}
\usepackage{amsmath}
\usepackage{amssymb}
\usepackage[utf8]{inputenc}
\usepackage{xcolor}
\usepackage{graphicx}
\usepackage{dcolumn}
\usepackage{bm}
\usepackage{mathtools,slashed}
\begin{document}

%\preprint{APS/123-QED}

%\title{Muon collider can be used as a Dark Matter detector}% Force 
\title{A new methodology for direct detection of heavy dark matter at intense particle beam facilities}
%line breaks with \\

\author[UoY]{A. Acar}
\author[UoY]{M.~Bashkanov}\ead{mikhail.bashkanov@york.ac.uk}
\author[UoY]{D.P. Watts}
\address[UoY]{School of Physics, Engineering and Technology, Department of Physics, University of York, Heslington, York, Y010 5DD, UK}
%\email{mikhail.bashkanov@york.ac.uk}

\date{\today}% It is always \today, today,

\begin{abstract}
We propose new concepts for experiments in which intense high energy photon or muon beams are employed parasitically to detect scattering by cosmic heavy weakly interacting dark matter (DM) particles. We show that the scattering cross-sections are sizeable enough to potentially observe beam scattering on heavy dark matter particles at high beam intensities for typically inferred near-Earth DM densities of $\rho_\chi\sim0.3~GeV/cm^3$. The predicted effect is particularly large in the case of a proposed muon collider Higgs factory, especially in the heavy (and poorly constrained) DM scenarios of WIMPZilla's. Current photon facilities such as at Jefferson Laboratory are predicted to require intensity and energy upgrades to reach detectable rates.
\end{abstract}

\maketitle

%\tableofcontents

\section{\label{sec:Intro} Introduction}
 Based on astronomical observations it is widely hypothesized that the ordinary luminous matter covers only 5\% of gravitating matter,
 and that the main contributor is invisible Dark Matter (DM), covering as much as 20\% of the energy balance of our Universe. The most prominent DM candidates currently are weakly interacting massive particles (WIMPs). These hypothetical particles are postulated not to interact strongly or electromagnetically, but only weakly and gravitationally \cite{WIMP1}. There are a lot of possible WIMP candidates with masses as light as a few GeV \cite{Baer:2014eja} up to as heavy as the Planck mass ($M_{Planck}$), the so called WIMPZillas \cite{Transplanckian}. Current constraints from underground DM search experiments limit the spin-independent DM-nucleon scattering cross section to be ($\sigma/M_{\chi}<3.7  \times 10^{-46}cm^2 \times {M_{\chi}}/{1 \\ \ \mathrm{TeV}}$, for $M_{\chi}>1$ TeV)~\cite{XENON:2025vwd}, where $M_{\chi}$ is the WIMP mass. 

 It was recently shown that the absence of a direct photon-DM coupling does not actually prevent photons from interacting with DM particles\footnote{ In the case of Beyond-Standard-Model (BSM) DM, one can have a direct coupling to a loop of SM particles without Higgs boson mediator $\chi\chi\to loop\to\gamma\gamma$ for the Majorana DM in a similar manner as $\chi\chi$ annihilation. However, such  photon-dark matter scattering was thought to give weak constraints, and often discarded as the cross section was too small \cite{McDermott_2011}.} ~\cite{RedorBlue}.  It was also demonstrated that the DM-photon interaction roughly scales quadratically with the energy of the photon~\cite{RedorBlue}, which creates an interesting opportunity to use beam-lines from high-energy, high-intensity photon or muon facilities as DM detectors. Not to produce them at these machines, but to detect existing cosmological DM.
 
\begin{figure}[!h]
\begin{center}
\includegraphics[width=0.47\textwidth,angle=0]{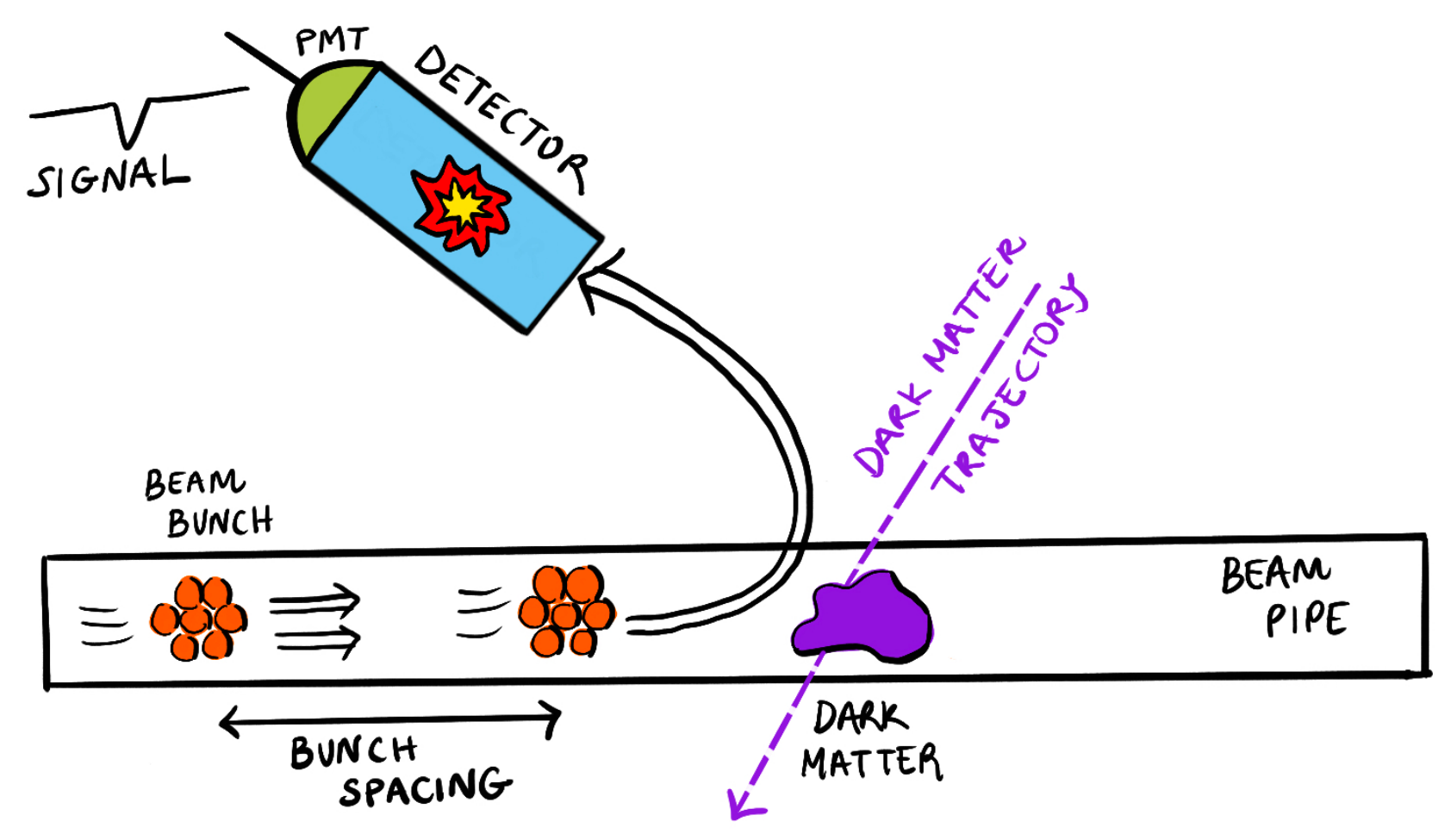}
\end{center}
\caption{Schematic illustration of the concept for detection of DM through scattering of particles from an intense particle beam.}
\label{experiment}
\end{figure}

 Intense high energy photon beams are mainly produced using  the bremsstrahlung of high energy electron beams on a thin radiator. The photons typically travel a significant distance in a vacuum beam pipe between the bremsstrahlung production site and the downstream experimental target. The distances are typically large. For example in the case of GlueX@JLab the distance between the bremsstrahlung radiator and the GlueX experiment is over 50 metres \cite{Adhikari:2020nima}. In this paper we estimate the probabilities  for the beam photons/muons to scatter on DM intercepting the beam. The basic principle is shown schematically in Fig.~\ref{experiment}. We use these calculations to infer what constraints are possible on  the maximum mass and the DM density, as a function of the achievable beam flux and using a range of DM masses.
 
 Beams at muon colliders offer complementary sensitivity through their direct interaction with DM via Higgs exchange (Fig~\ref{Diagrams} (c)) \footnote{It is assumed that dark matter has a very small or even zero weak charge, so direct interactions with other weak bosons are discarded}. Although the Higgs coupling with the muon is smaller than in the photon case, the gains in beam energy and luminosity give clear prospects. The coupling to the muon is weaker as the indirect Higgs boson to $\gamma\gamma$ coupling (width) is around 9.2~keV, an order of magnitude larger than the direct $\Gamma(H\rightarrow \mu^+\mu^-)=0.91$~keV (the photon coupling is enhanced due to the Higgs-photon interactions via the $W^{\pm}$ and $t-$quark loops; see Fig.~\ref{Diagrams}(a,b). 
 
 %using asymmetries in the yields from positively and negatively charged muons in the ring opens up another interesting field of study. 

% In order for the Universe to exist as a matter over antimatter dominated object, a sizeable CP- violation is needed to fulfil the Sakharov criteria \cite{Sakharov1967}. Years of research of CP violating processes in strange, charm, beauty and neutrino sectors did not give us any results on where this large CP violation is originating from. Based on the current Standard Model the universe should not exist, but it does. One can assume that part of the necessary CP violation possibly originates from the dark sector. In this case the cross-section for scattering of particles and anti-particles over DM could be different, meaning that there will be more $\mu^+ \chi$ scattered events than $\mu^- \chi$ or vice versa. Having identical $\mu^+/\mu^-$ fluxes, the same location and length of the "target" we can test this hypothesis with unprecedented systematic accuracy. One should note that such CP-violation in the dark sector would necessitate other diagrams besides Higgs exchange to make it detectable, and in general would likely require other BSM interactions, leaving such evaluations outside the scope of this paper using the polarisation of the beams. 

 %In the model of \cite{RedorBlue}, the DM particles interact with the normal world only via the Higgs boson exchange   And since the Higgs boson to $\gamma\gamma$ coupling is large, the $\Gamma(H\rightarrow \gamma\gamma)=9.2$~keV width is an order of magnitude larger than the $\Gamma(H\rightarrow \mu^+\mu^-)=0.91$~keV, due to Higgs-photon interactions via the $W^{\pm}$ and $t-$quark loops, see Fig.~\ref{Diagrams}, photon-DM scattering cross-section can be sizeable. 

 The use of muon beams as a DM detector also has distinct advantages from technical considerations. Despite their smaller coupling to Higgs, the higher energies and luminosities provide opportunities for more stringent constraints. At the proposed future muon collider facility, the energy of the muon would be $E_\mu=M_H/2$ - half of the Higgs mass and significantly higher than achievable at current (or proposed) photon facilities. The number of muons in the beam, $2\times 10^{12}$, and an expected  400 km path (from recirculation around the 10 km storage ring), as will be shown, give significant enhancements in the detection probabilities. Longer term, if muon DM interactions are observed there are prospects for future study of CP-symmetry breaking in the dark sector, which would reveal itself as asymmetries in the yields of $\mu^+ \chi$ and $\mu^- \chi$ scatter events\footnote{Note that Higgs exchange by itself would not lead to any CP-violating asymmetries in $\mu^+/\mu^-$}. "Non-dark" strange, charm, beauty and neutrino sectors give so far null results despite larger than SM-model CP violation being crucial to explain the matter dominated universe and to fulfil the Sakharov criteria \cite{Sakharov1967}\footnote{Such CP-violation estimates in the dark sector would necessitate other diagrams besides Higgs exchange to make it detectable, likely requiring BSM interactions, and are outside of the scope of current models presented here}.  
 
 Current photon beams have lower intensity, path lengths (beam recirculation is not possible) and are not mono-energetic. Despite these limitations we will show that upgraded facilities (e.g. at Jefferson Lab) may also offer prospects to provide DM constraints with the new method. 
 
 %: majority of photons have low energy due to bremsstrahlung production. Although the length of the target can be large $\sim 100m$, it is still finite. On the other hand, muons, despite having a smaller coupling to Higgs, can have really high energy and luminosity. At a future muon collider facility, the energy of the muon would be $E_\mu=M_H/2$ half of the Higgs mass - much higher than the energy of the photon at any existing photon facility. The number of muons in the beam, $2\times 10^{12}$, will also be very large and moving in circles over the 10~km long ring. They will travel for nearly 400 km before being replenished!

In this paper we calculate the cross section and event yields for photon and muon beams scattering off DM particles for various existing and future facilities, calculated using the theoretical ansatz presented  in Ref.~\cite{RedorBlue}. %While current facilities offer very limited possibilities to constrain the DM mass/density, next generation intense facilities utilizing photon or muon beams 
The prospect to substantially restrict the available parameter space for existing DM models in this poorly constrained regime of heavy WIMP DM is indicated. 

The paper is organized as follows: in Section II we introduce possible weak light-DM and muon-DM scattering processes and calculate their probabilities; in Section III we show results, using these cross sections to estimate the event yields expected for existing and planned beam facilities.

\section{Weak Interactions}
\begin{figure}[!h]
\begin{center}
\includegraphics[width=0.48\textwidth,angle=0]{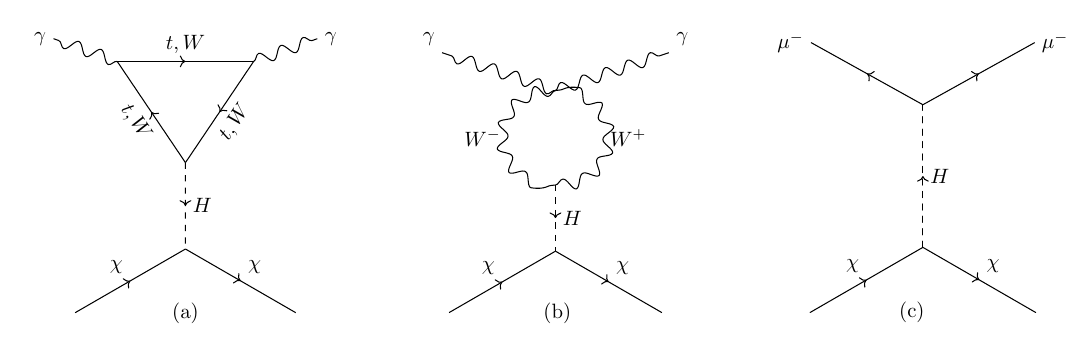}
\end{center}
\caption{All possible lowest order diagrams for dark matter-photon and DM-muon scattering propagated by the Higgs boson in the unitary gauge.}
\label{Diagrams}
\end{figure}

Since the introduction of weak interaction theory by Weinberg and Salam \cite{Weinberg} possible extensions to the Standard Model have been hypothesized.
%One of the appealing possibilities, the supersymmetry (SUSY) model might provide a reasonable DM candidate, neutralino, and also solve all renormalisation issues at high energies. Unfortunately, the LHC has found no signs of SUSY particles as yet. The discovery of a Higgs boson with rather large mass, $M_H=125$~GeV \cite{ATLAS}, and consistent standard model decay branches \cite{Arbey:2011ab} also tends to disfavour SUSY-theories in the sensitivity ranges currently accessible in experiment.
In recent years,  many experiments have aimed to discover BSM physics ~\cite{FASER}, but Standard Model (SM) violations have not been discovered so far. In our calculations (based on the work presented in Ref.~\cite{RedorBlue}) we assume that any heavy weakly interacting DM particle interacts within the framework of the established SM, without introduction of any additional forces, fields or conservation laws. In this paradigm one can still introduce a WIMP DM particle, {\it e.g.} in the form of a very heavy 4th generation neutrino \cite{DarkNeutrino}, a right-handed neutrino \cite{righthanded} or, indeed, any stable fermion. In any of these cases a fermionic DM particle would necessarily get its mass through the Yukawa coupling to the Higgs boson. In this scenario, DM would unavoidably interact with light, since DM must interact with the Higgs boson, and the well-measured Higgs decay into two photons tells us that a photon can scatter off of a dark matter particle mediated by the Higgs boson. On a diagrammatic level, this process is expressed in Fig.~\ref{Diagrams} (a) and (b), in the unitary gauge~\footnote{The calculations in this work are performed in the unitary gauge to avoid ghost and Goldstone boson loops.}.  

\subsection{Photon-Dark Matter Scattering}
Following the approach of $H\rightarrow\gamma\gamma$ calculations from Ref.~\cite{Marciano}, the DM-photon scattering matrix element can be expressed as:

\begin{align}
i\mathcal{M} = i\mathcal{M_F} + i\mathcal{M_{W}}
\end{align}

\noindent where $i\mathcal{M_F}$ and $i\mathcal{M_{W}}$ are the matrix elements of processes involving the fermion and W-loop respectively. Only the top quark is considered for the fermion loop, as the other particles are too light to contribute significantly to the cross section. Details of the calculations can be found in Ref.~\cite{RedorBlue}.
% The full matrix element integrated over Feynman parameters after dimensional regularization is given by: 

% \begin{align}
% \nonumber i\mathcal{M} = \frac{\alpha g_w}{4\pi m_w} ((k_1 \cdot k_2)g^{\mu\nu} - k_1^\nu k_2^\mu) \epsilon^*_\mu (k_1) \epsilon_\nu (k_2) \\
% \cdot \frac{1}{t-m_H^2+im_H\Gamma_H} [\bar{u}(p_2)\frac{g_w}{2m_w}m_{\chi} u(p_1)] [N_c Q_f^2 I_F + I_W]
% \end{align}

% %% This will need to be rewritten.
% \noindent where $\alpha$ is the fine structure constant, $g_w$ is the weak W-boson coupling constant and $m_w$ is the mass of the W-boson. $k_1$, $\epsilon_\mu$ and $k_2$, $\epsilon_\nu$ are the 4-vectors and polarization vectors of outgoing and incoming photons respectively. $m_H$ and  $\Gamma_H$ are the mass and the width of Higgs, $N_c$ is the number of quark colours, $Q_f$ is the charge of the quark depending on its flavour and $m_{\chi}$ is the mass of the dark matter particle. $t$ has the standard definition of the Mandelstam variable, and the $u$ stands for a Dirac spinor. The terms $I_F$ and $I_W$, which come from Feynman parameter integrals, are given by:

% \begin{align}
% I_F(\beta)= -2\beta(1+ (1-\beta)f(\beta))
% \\
% I_W(\beta)= 2 + 3\beta+3(2\beta-\beta^2)f(\beta)
% \end{align}

% where $f(\beta)$ is defined as:

% \begin{align}
% f(\beta)=
% \begin{cases}
%   \arcsin^2(\beta^{-1/2}), & \text{if } \beta \ge 1,\\[1ex]
%   -\frac{1}{4} [\ln\frac{1+ \sqrt{1-\beta}}{1-\sqrt{1-\beta}} -i\pi]^2,  & \text{if } \beta < 1.
% \end{cases}
% \end{align}

% $\beta$ is $\beta=-4m_w^2/t$ for the W-boson loop, and $\beta=-4m_t^2/t$ for the top quark loop.

\noindent From the square of the full matrix element~\footnote{Since the Higgs is a scalar particle, there is no polarization information transferred: we can expect the $H \chi \chi$ vertex to decouple from the $H\gamma\gamma$ vertex, which makes it easier to obtain the square of the matrix element.}, one can obtain the differential cross section by multiplying it by the relevant flux and phase-space terms\footnote{Full derivation of scattering cross-section calculations can be found in supplementary materials of Ref.~\cite{RedorBlue}}:

\begin{align}
\frac{d \sigma}{d \Omega} = \frac{\alpha^2 g_w^4}{(4\pi)^2 m_w^4} \frac{3t^2}{8} \frac{m_\chi^2 (2m_\chi^2 - \frac{t}{2})}{(t-m_H^2)^2 + m_H^2 \Gamma_H^2} \frac {\abs{I_w + N_c Q_f^2 I_f}^2} {64\pi^2 s}
\end{align}

where $\alpha$ is the fine structure constant, $g_w$ is the weak W-boson coupling constant and $m_w$ is the mass of the W-boson. $k_1$, $\epsilon_\mu$ and $k_2$, $\epsilon_\nu$ are the 4-vectors and polarization vectors of the outgoing and incoming photons, respectively. $m_H$ and $\Gamma_H$ are the mass and width of Higgs, $N_c$ is the number of quark colors, $Q_f$ is the charge of the quark depending on its flavor, and $m_{\chi}$ is the mass of the dark matter particle. $t$ has the standard definition of the Mandelstam variable and $u$ stands for a Dirac spinor. The terms $I_F$ and $I_W$, which come from Feynman parameter integrals, are given by:

 \begin{align}
 I_F(\beta)= -2\beta(1+ (1-\beta)f(\beta))
 \\
 I_W(\beta)= 2 + 3\beta+3(2\beta-\beta^2)f(\beta)
 \end{align}

 where $f(\beta)$ is defined as:

 \begin{align}
 f(\beta)=
 \begin{cases}
   \arcsin^2(\beta^{-1/2}), & \text{if } \beta \ge 1,\\[1ex]
   -\frac{1}{4} [\ln\frac{1+ \sqrt{1-\beta}}{1-\sqrt{1-\beta}} -i\pi]^2,  & \text{if } \beta < 1.
 \end{cases}
 \end{align}

 $\beta$ is $\beta=-4m_w^2/t$ for the W-boson loop and $\beta=-4m_t^2/t$ for the top quark loop.

\subsection{Muon-Dark Matter Scattering}

The DM-muon scattering, $\mu (p_1) \chi (p_2) \rightarrow \mu (p_3) \chi (p_4) $,  matrix element  is given by:

\begin{align}
\nonumber i\mathcal{M} = \frac{-i}{t-m_H^2+im_H\Gamma_H} \left[\bar{u}(p_3)\frac{g_w}{2m_w}m_\mu u(p_1)\right]  \left[\bar{u}(p_4)\frac{g_w}{2m_w} m_{\chi}u(p_2) \right] 
\end{align}

The differential cross section in the CMS frame is given by:

\begin{align}
\frac{d \sigma}{d \Omega}_{CMS} &= \frac{\mathcal{M}^2}{64 \pi^2 s} \nonumber
\end{align}
where {\it s} is the Mandelstam variable, and $\mathcal{M}^2$ is given by:
\begin{align}
\mathcal{M}^2= \frac{m_\mu^2m_{\chi}^2}{(t-m_H^2)^2+ m_H^2 \Gamma_H^2} \left(\frac{4m_\mu^2 -t}{2} \right) \left( \frac{4m_\chi^2 -t}{2} \right) \frac{g_w^4}{m_w^4}
\end{align}
The differential cross section in the lab frame is given by:

\begin{align}
\frac{d \sigma}{d \Omega}_{lab} &= \frac{1}{64 \pi^2} \frac{1}{\abs{\vec{p_1}} m_{\chi}} \frac{\abs{\vec{p_3}}^2}{\abs{\vec{p_3}}(E_1 + m_{\chi}) - E_3\abs{\vec{p_1} \mathrm{cos(\theta)}}}  \mathcal{M}^2
\end{align}

where $|\vec{p_1}|$ and $|\vec{p_3}|$ are the absolute value of the 3-momentum of the incoming and scattered muon respectively, $E_1$ and $E_3$ are their energies, and $\theta$ is the angle between them, all given in the laboratory frame.

\section{Results}
In this section the possibility for a range of photonic and muonic beams to scatter off  DM will be estimated, accounting for luminosity, path length and DM mass. Firstly, photon beams from laser- plasma accelerators, such as the RAL facility, will be examined \cite{Kettle2021LaserPlasma}. Subsequently estimates from conventional photon machines such as the Compact Photon Source (CPS) at Jefferson Lab (JLab), Hall-D for KLF experiment \cite{Amaryan2020KLBeam} and future JLab-CPS upgrades will be discussed. Finally, the prospects for the proposed muon collider \cite{InternationalMuonCollider:2025sys} will be presented. For all of these possibilities, a range of DM densities from $\rho_\chi\in[0.01,1000]~GeV/cm^3$ will be employed\footnote{Note that a standard DM density in the vicinity of earth is assumed to be $\rho_\chi\sim 0.03~GeV/cm^3$, while the DM density in the galaxy centre and in DM clumps can reach $\rho_\chi\sim 1000~GeV/cm^3$, \cite{Riccardo_Catena_2010, Arbuzova2024CosmicRays}}.

\begin{figure}[!h]
\begin{center}
\includegraphics[width=0.35\textwidth,angle=0]{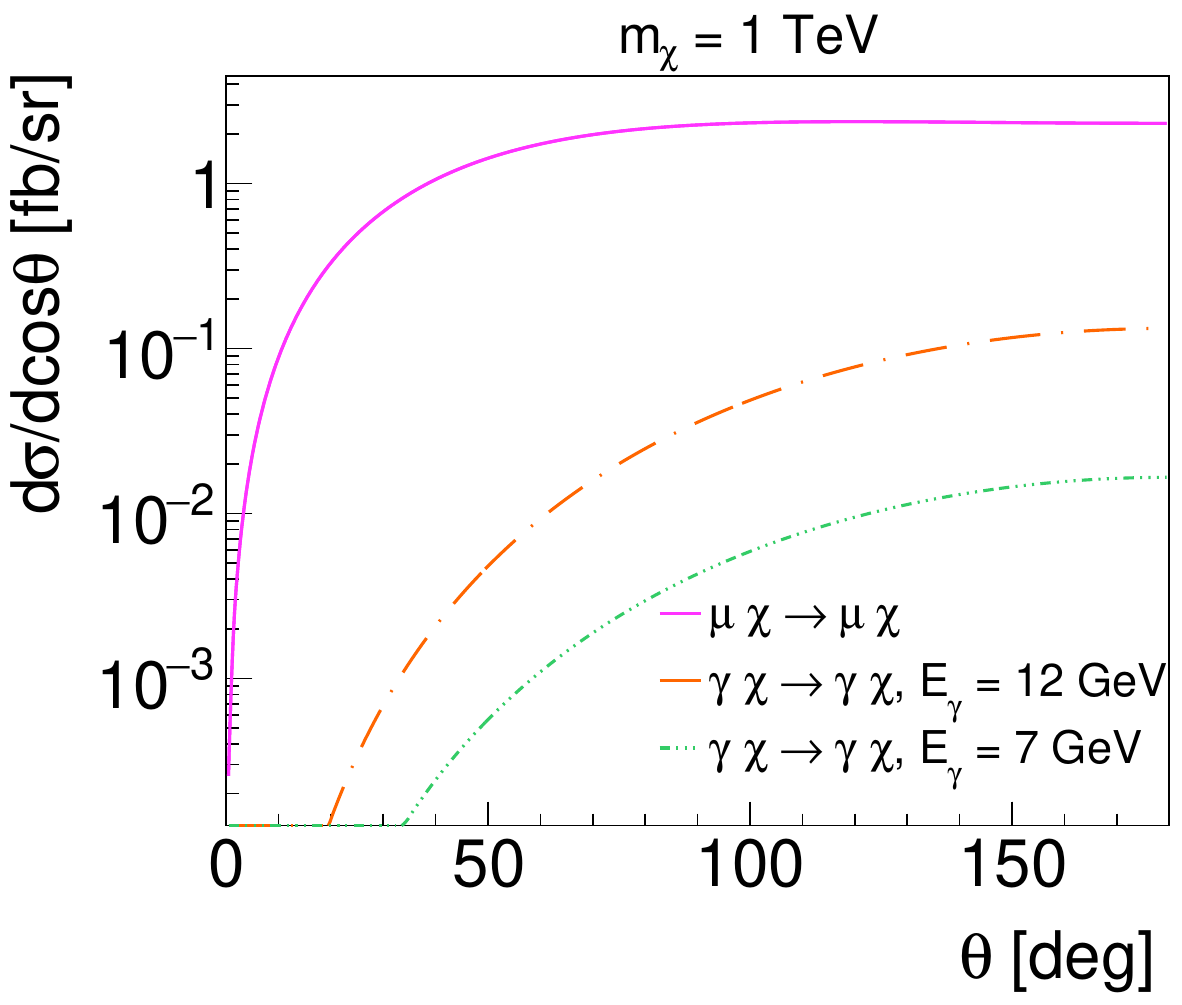}
\end{center}
\caption{ Differential cross-sections for $E_{\mu}=M_{H}/2$ muons (pink), 12~GeV photons(dash-dot orange) and 7~GeV photons (dash-dot-dot green) scattered on 1 TeV WIMP dark matter particles as a function of photon scattering angle in the lab frame.}
\label{diff}
\end{figure}

The calculated DM scattering cross sections as a function of scatter angle for a 1 TeV DM mass and for muonic and photonic scattering are shown in Fig.~\ref{diff}. This figure illustrates a few general principles for achieving sensitivity in experiments, which we discuss before more detailed discussion in the following sections. Firstly, the cross section is much larger at backward scatter angles so these regions offer the best prospects for detection. As the DM mass is large these backward scattered particles will be close to 100\% of the incident beam energy, giving a strong signature for scattering off a heavy WIMP. Scattering from background gas nuclei in the beam line would inevitably lead to a diminishing of the backscattered energy due to the higher energy taken by the (relatively light) nuclear recoil. For example in the case of a high-energy photon, ~60 GeV, scattering on a nucleon with a scattering angle as high as 175 degrees, its energy would only be 80\% of the incoming beam energy. The role of such background processes will also be constrained by the bunched nature of the beams. For example, at KLF@JLab beam bunches will be spaced by 64~ns while at laser-plasma accelerators the repetition rate is even lower - 1Hz-1kHz. Timing correlations between the beam-bunch and hit in the detector would enable veto of any high energy scattered particles upstream of the detector and provide spatial localisation of downstream scatter sites. One can further enforce this constraint if the detector is able to map the photon shower profile for directionality to the interaction point, such as achievable with highly segmented calorimeters like those being developed for the  EIC spaghetti calorimeter \cite{BNL68933_2022}. 

It is also important to note how the DM scattering cross section is strongly enhanced with increasing beam energy. For the case of bremsstrahlung photon beams, which have a broad energy spread up to the electron beam energy, the probability of DM scattering dominantly occur for the higher energy part of the bremsstrahlung flux, while for background processes larger event rates would typically be present for the lower energy parts of the flux. Muonic beams can be monoenergetic, so would be expected to have smaller contributions from such types of background processes.

%the scattered photon spectrum would be observed, Fig.~\ref{phot_flux}(dashed), while for the background all photons, predominantly the low energy ones, would contribute. Also, if the scattering occurs on a very heavy dark matter particle, the energy of the scattered photon/muon would not change, while if it scatters on a light nucleon of residual beam gas, some energy will be spent on nuclear recoil. If a high-energy photon, ~60 GeV, scatters on a nucleon with a scattering angle as high as 175 degrees, its energy would only be 80\% of the incoming beam energy, while if it scatters on a DM particle, its energy would be close to 100\%. These features - timing, energy, and uniformity of DM over the beam path should allow one to disentangle the DM scattering from conventional background. 

\begin{figure}[!h]
\begin{center}
\includegraphics[width=0.40\textwidth,angle=0]{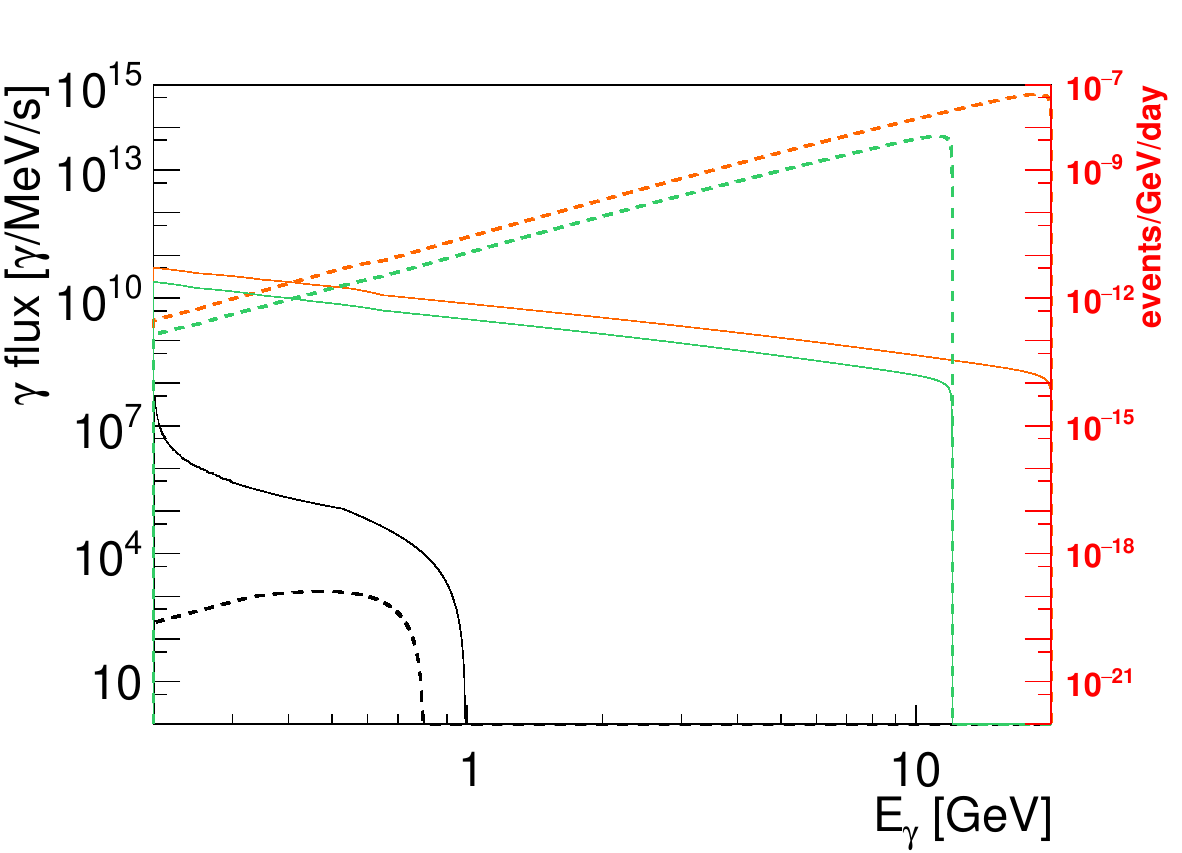}
\end{center}
\caption{Energy dependence of the photon flux(solid, left axis) and a corresponding number of scattered photons(dashed, right axis) for RAL(black), 12~GeV JLab(green) and 20~GeV JLab(orange). $M_\chi=M_{Planck}$. $I_{JLab}^{e^{-}}=5\mu A$.}
\label{phot_flux}
\end{figure}

\subsection{Laser-plasma accelerators}
Laser plasma accelerators offer extremely high photon fluxes ($\sim 10^9~\gamma$'s per shot). Photon beams can be transferred for many meters in vacuum making them a viable option to consider as a DM detector via parasitic $\gamma-\chi$ scattering.
Let's consider the simple case if the photon-DM scattering can be detected at the existing RAL beam-line \cite{Kettle_2021} \cite{Kettle2021LaserPlasma}. If we assume $\sim 10^9$ photons per shot, the photon spectrum from \cite{Kettle2021LaserPlasma}, 1 m beam-line distance, with an optimistic 1 Hz shot rate we expect  $\sim 10^{-20}$ scatterings a day for the realistic $\rho_{\chi}=0.3~GeV/cm^3$ density, and $M_\chi=M_{Planck}$, see Fig~\ref{phot_flux}. The extremely low probability of detection makes this a very unviable experiment. However, if one would be able to further increase shot frequency and electron energy, such an experiment might become feasible. As one can see from Eq.~2 and Fig.~\ref{diff}, due to the quadratic rise of the $\gamma\chi$ cross-section with photon energy, the spectrum of scattered photons would be much harder and very different from both the incoming photon spectrum and the background Compton spectra.  

\subsection{Conventional photon beams}

While conventional photon beams typically have much smaller fluences per shot compared to laser-plasma accelerators, the overall photon flux exceeds that currently available from laser accelerators. The KLF experiment at JLab can be considered as an example, where a $5\mu A$ electron beam with 64ns bunch spacing and  $\sim 50$ m beam pipe leads to production of $4.7\cdot10^{12}$ photons with $E_{\gamma}>1.5$~GeV \cite{Amaryan2020KLBeam}, with a photon spectrum extending as high as 12 GeV using a compact photon source (CPS), see Fig.~\ref{phot_flux}. The bremsstrahlung photon spectrum falls sharply with energy, but the cross-section rises similarly fast, leading to a peak in event rate close to a maximum accessible energy. A 64 ns spacing is equal to nearly 20 m distance between bunches, leading to good temporal separation of potential scattered photons and beam induced background. For 200 days beam-time, that would lead to $5\cdot 10^{-6}$ events of $\gamma\chi$ scattering. 
%, which is a lot more favorable compared to $\sim 10^{-20}$ events for the RAL case. 

One needs to note that $5\mu A$ current, which defines photon flux, originates not from an accelerator restriction, but from the CPS shielding mass, which should not exceed 100 tons to cope with the concrete floor strength. The CEBAF accelerator can deliver as high as 50mA electron current \cite{Reece2016} and the radiator thickness can be increased from 10\%~RL (Radiation Length) to 20\%~RL, leading to about 0.1 $\gamma\chi$ scattering per experiment for the realistic $\rho_{\chi}=0.3~GeV/cm^3$ density and a $M_\chi=M_{Planck}$. If a DM filament or clump with assumed $\rho_{\chi}=1000~GeV/cm^3$ encounters such a beam, the scatter rate is estimated to reachnearly one event per hour. 

\begin{figure}[!h]
\begin{center}
\includegraphics[width=0.35\textwidth,angle=0]{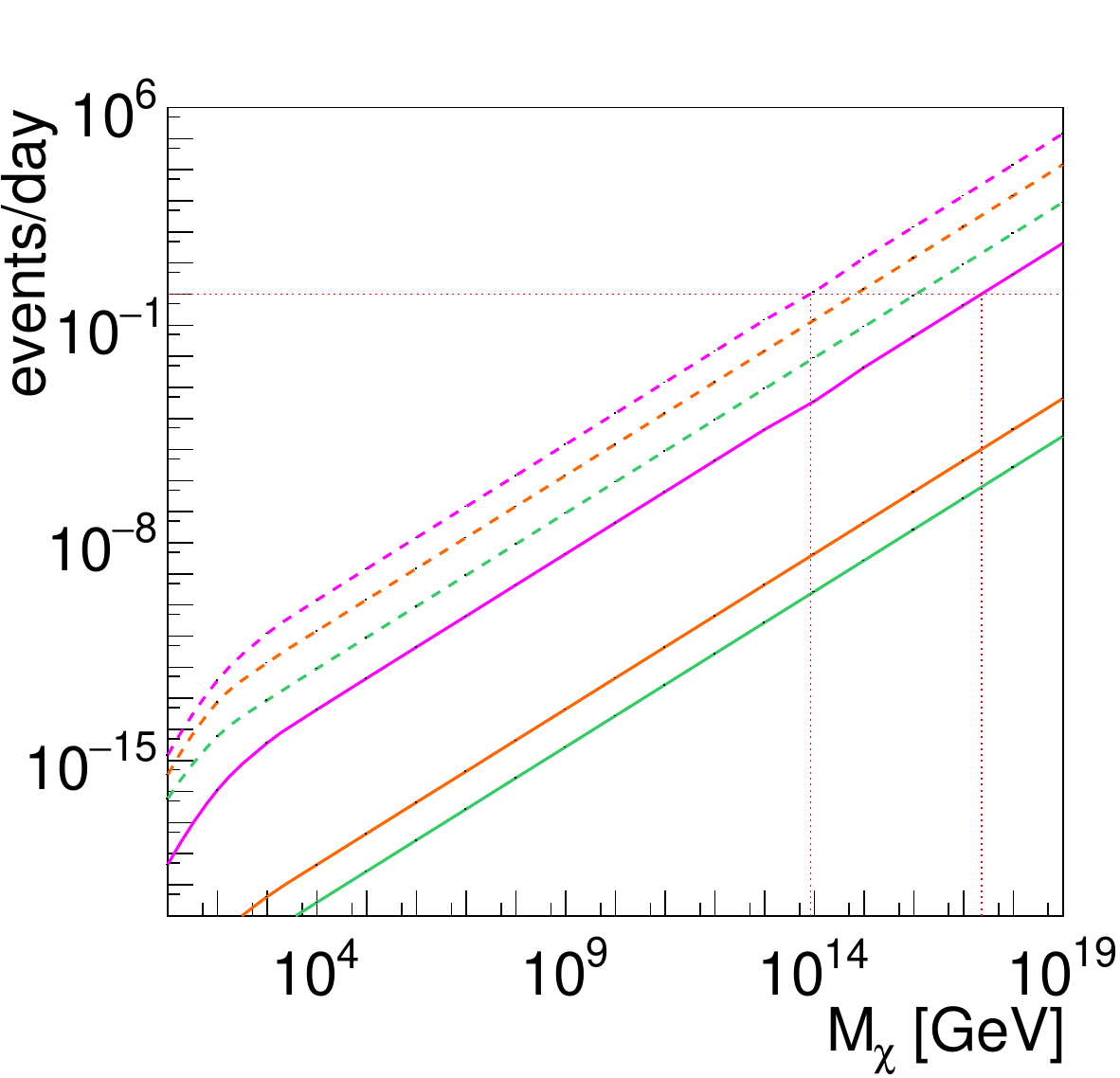}
\end{center}
\caption{Rate dependence of the beam scattering events as a function of DM mass, $M_{\chi}$ for muon collider (pink), 12~GeV JLab (green) and 20~GeV JLab(orange). Solid -  for the normal conditions ($\rho_{\chi}=0.3~GeV/cm^3$, $I_{JLab}^{e^{-}}=5\mu A$, 10\%~RL radiator). Dashed for "lucky" conditions ($\rho_{\chi}=1000~GeV/cm^3$, $I_{JLab}^{e^{-}}=50mA$). Dotted red lines highlight 1 event per day case.}
\label{DM_rate}
\end{figure}

There is a proposal to extend the CEBAF electron beam energy to 20-22 GeV. If one takes a typical bremsstrahlung spectrum for 20 GeV electrons, 50 mA beam current, assume the same  20\%~RL CPS and $\sim 50$ m long beam-line length, $\rho_{\chi}=1000~ GeV/cm^3$, a very impressive rate of an event occurring every couple of seconds can be reached, see Fig.~\ref{DM_rate}. Such rates might provide sensitivities to DM as low as $M_{\chi}\sim 10^{13}$~GeV.

\subsection{Muon collider}

There is another interesting opportunity to look for high-energy beam scattering on ambient DM - a muon collider. From Section~3 one can see that the muon-DM scattering cross-section is somewhat smaller than the photon-DM case for the same energies. However, this is offset by the monoenergetic nature of such beams and the higher luminosities available. The bremsstrahlung origin of photon beams leads to the unfavourable situation where most of the flux is concentrated at small energies, and the useful high energy portion of the flux is relatively small. The (charged) muon beams also have the advantage that they can be recirculated offering greatly enhanced path lengths for interaction. In detection of backscattered particles, muons have additional advantages over photons. Reconstructed muon tracks can give the position of scattering with higher ($\sim$mm) precision and timing to $\sim$ps accuracy.  The main purpose of a future muon collider is to build a Higgs factory, meaning that muons will have energies equal to half of the Higgs mass $E_\mu=M_H/2\sim 63$~GeV - much larger than possible with current or planned photon beam technologies.  
%have another advantage over photons - they are charged: photons only travel straight, so after they fly the 50~m long beam-line, they are gone. Muons in a muon collider, on the other hand, travel in circles, so the same muon can be re-utilized multiple times. Also, the circumference of the muon ring is about 10~km - much larger than anything available for photon beams. 
These advantages significantly increase the rates for the $\mu-\chi$ scattering at a muon collider to a sizeable $\sim 1$ event per hour for $\rho_{\chi}=0.3~GeV/cm^3$, $M_\chi=M_{Planck}$ and $2\times 10^{12}$ muons in the beam, Fig.~\ref{DM_rate}. A local increase in DM density can lead to enhanced rates of scattering.

%Like photons, muons prefer to scatter on DM particles in a backward direction, simplifying the background suppression. A mono-energetic spectrum of muons in a muon collider further simplify the background suppression. 

After the main "Higgs factory" program, the muon collider is expected to operate at higher energies, up to $\sqrt{s}\sim 10$~TeV and possibly even higher luminosities than $2\times 10^{12}$ particles in the ring \cite{DelBlas2023MuonCollider}. This would increase the predicted rate of $\mu\chi$ scattering very significantly, allowing to constrain the DM mass-cross-section dependence to lower mass values. 

\section{\label{sec:final} Summary}

%Particle experiments are known to be very sensitive: LHC can trace the motion of the Moon, neutrino experiments at Gran Sasso can see surface temperature and precipitation despite 3.5 km of water equivalence rock on top. So it is not a surprise that modern experiments can be parasitically used to measure something different to what they were originally designed for. 
In this paper we have shown that future high intensity, high energy particle beams can potentially be utilized to constrain the existence and properties of heavy weakly interacting particles in the WIMPZilla domain. %Sources of WIMPZillas include a heavy fourth generation neutrino.
The sensitivity is estimated using calculations of the interaction of beam particles with ambient DM in the beamline assuming interactions mediated by Higgs exchange and taking facilities at the frontiers of luminosity. The prospects for constraining this heavy dark sector in direct experiments is assessed for the first time, establishing a new methodology which could significantly extend the sensitive mass ranges of terrestrial DM experiments. The proposed future muon collider is predicted to give measurable event rates of elastically backscattered beam particles of around 1 event per hour for realistic (average) DM densities at the earth's surface and for scattering from Planck mass scale WIMPZilla's. Current intense photon beam facilities would require upgrades in energy and luminosity to provide meaningful constraints.

\section{Acknowledgements}

We want to thank Steven J Rose, Stuart P D Mangles and Brendan Kettle from Imperial College London for fruitful discussions.
This work has been supported by the U.K. STFC (ST/L00478X/1, ST/T002077/1, ST/L005824/1, 57071/1, 50727/1 ) grants. After publication all data will be available at Pure@York~\cite{yorkdataset2026}.
%************
%\section{Acknowledgement} 

\bibliographystyle{elsarticle-num}   % or another journal-required style
\bibliography{references_v3}

\end{document}